\documentclass[epj,final,amsmath,amssymb,pdftex]{svjour}
\pdfoutput=1

\def\<{{<}}
\def\>{{>}}

\usepackage{graphicx}
\usepackage{amssymb}

\input{epsf}

\begin{document}

\title{Poincar\'e recurrences and Ulam method
\\ for the Chirikov standard map}

\author{Klaus M. Frahm and Dima L. Shepelyansky}

\institute{
Laboratoire de Physique Th\'eorique du CNRS, IRSAMC, 
Universit\'e de Toulouse, UPS, 31062 Toulouse, France
\and
http://www.quantware.ups-tlse.fr}

\titlerunning{Poincar\'e recurrences for the Chirikov standard map}
\authorrunning{K.M.Frahm and D.L.Shepelyansky}

\date{February 12, 2013}

\abstract{We study numerically the statistics of Poincar\'e
recurrences for the Chirikov standard map and the separatrix map
at parameters with a critical golden invariant curve.
The properties of recurrences are analyzed with 
the help of a generalized Ulam method.
This method allows to construct the corresponding Ulam matrix
whose spectrum and eigenstates are analyzed
by the powerful Arnoldi method.
We also develop a new survival Monte Carlo method
which allows us to study recurrences 
on times changing by ten orders of magnitude.
We show that the recurrences at long times are determined
by trajectory sticking in a vicinity of the critical golden curve
and secondary resonance structures.
The values of Poincar\'e exponents of recurrences are  
determined for the two maps studied.
We also discuss the localization properties of eigenstates
of the Ulam matrix and their relation 
with the Poincar\'e recurrences. 
}

\maketitle
\section{Introduction}
\label{sec1}

The interest to understanding of transition from dynamical 
to statistical description of motion had 
started from the dispute between Loschmidt and Boltzmann,
which is now known 
as the Loschmidt paradox {\cite{loschmidt,boltzmann}.
The  two-dimensional (2D) symplectic maps represent an excellent laboratory
for  investigation of how statistical laws
appear in dynamical, fully deterministic systems.
Their properties have been studied in great detail during
last decades both on mathematical 
(see e.g. \cite{arnold,sinai} and Refs. therein)
and physical (see e.g. \cite{chirikov1969,chirikov1979,lichtenberg} 
and Refs. therein) levels of rigor.
The case of completely chaotic behavior, appearing  e.g. in Anosov systems,
is now well understood \cite{arnold,sinai} but
a generic case of  maps with
divided phase space, where islands of stability
are surrounded by chaotic  components, still preserves 
its puzzles. A typical example of such 
a map is the Chirikov standard map \cite{chirikov1969,chirikov1979}
which often gives a local description of dynamical chaos
in other dynamical maps and describes
a variety of physical systems (see e.g. \cite{scholar}).
This map has the form:
\begin{equation}
\label{eq_stmap}
{\bar y} = y + \frac{K}{2\pi} \sin (2\pi x) \; , \;\; 
{\bar x} = x + {\bar y} \;\; ({\rm mod} \; 1) \;.
\end{equation}
Here $x,y$ are canonical conjugated variables of generalized phase and action,
bars mark the variables after one map iteration
and we consider the dynamics to be periodic on  a torus so that
$0 \leq x \leq 1$, $0 \leq y \leq 1$. The dynamics is characterized by
one dimensionless chaos parameter $K$.

For small values of $K$ the phase space  is covered by invariant
Kolmogorov-Arnold-Moser (KAM) curves which restrict
dynamics in the action variable $y$. For $K>K_g$
the last invariant golden curve with the rotation
number $r=r_g=\langle(x_t-x_0)/t\rangle=(\sqrt{5}-1)/2$
is destroyed \cite{greene,mackay} and it 
is believed that for $K>K_g$
the dynamics in $y$ becomes 
unbounded \cite{percival,chirikov2000}.
A renormalization technique
developed by Greene and MacKay \cite{greene,mackay} allowed 
to determine $K_g=0.971635406$
with enormous precision. The properties of the critical golden curve
on small scales are universal for all critical curves with the 
golden tail of the continuous fraction expansion of $r$
for all smooth 2D symplectic maps \cite{mackay}.
Here and below the time $t$ is measured in number of map iterations
(due to symmetry there is also 
a symmetric critical curve at $r=1-r_g$ at $K_g$).
For $K>K_g$ the golden KAM curve is replaced 
by a cantori \cite{aubry}
which can significantly affect the diffusive transport
through the chaotic part of the phase space \cite{meiss,stark}.
At any $K$ there are some chaotic regions in the phase space 
bounded by internal or isolating (at $K<K_g$)
invariant curves.

The dynamics inside a chaotic component of the phase space $(x,y)$
is characterized by correlation functions whose decay ensures
a transition from dynamical to statistical description.
The decay of correlations is related to the probability
to stay in a given region of phase space since 
for a trajectory remaining in a small region
the dynamical variables are strongly correlated.
This probability in its own turn
is related to the statistics of Poincar\'e recurrences.
Indeed, according to the Poincar\'e recurrence theorem
\cite{poincare}  a dynamical trajectory of a Hamiltonian system with
bounded phase space  always returns, after a certain time,
to a close vicinity of an initial state. 
However, the statistics of these recurrences 
depends on dynamical properties of the system.
For a fully chaotic phase space a probability to stay 
in a certain part of a phase space decays 
exponentially with time being similar to
a random coin flipping \cite{arnold,sinai}.
However, in dynamical maps with divided phase space,
like the Chirikov standard map, the decay 
of probability of Poincar\'e recurrences $P(t)$
is characterized by a power law decay
$P(t) \propto 1/t^\beta$ has $\beta \approx 1.5$
whose properties still remain poorly 
understood.

One of the first studies of Poincar\'e recurrences 
in dynamical Hamiltonian systems
with two degrees of freedom was done in 
\cite{lebowitz} where an algebraic decay with an
exponent $\beta=1/2$ was found. This exponent corresponds to
an unlimited diffusion on an infinite one-dimesional line
which is in contrast to a bounded phase space.
This strange observation was explained in \cite{kiev,kievb}
as a diffusion in a chaotic separatrix layer of a 
nonlinear resonance which takes place on 
relatively short diffusion times.
On larger times, which were not accessible to the computations
presented in \cite{lebowitz},
this diffusion becomes bounded by a finite width of the 
separatrix layer and a universal algebraic decay
takes place with the exponent $\beta \approx 1.5$
corresponding to a finite chaos measure  \cite{kiev,kievb}.
This algebraic decay of $P(t)$
has been confirmed by various groups in various
Hamiltonian systems \cite{karney,chsh},\cite{ott},
\cite{chirikov1999,chirikov1999b},\cite{ketzmerickcom,ketzmerickpre},
\cite{ketzmerick,artuso},\cite{venegeroles,shevchenko}.

One can argue that such a slow algebraic decay with $\beta \approx 1.5$
appears due to trajectory
sticking near stable islands and critical invariant 
curves and leads to an even slower correlation function decay
$C(t) \sim t P(t)$
with a divergence of certain second moments. 
A sticking in a vicinity of the critical golden curve \cite{mackay}
is expected to give $\beta \approx 3$ \cite{chirikov1999,chirikov1999b},
being significantly larger
than the average value $\beta \approx 1.5$.
A certain numerical evidence is presented in \cite{ketzmerickpre}
showing that long time sticking orbits
can be trapped not only in a vicinity
of a critical golden curve but also
in internal chaotic layers of secondary resonances. 

Theoretical attempts to describe trapping in secondary
resonances as renormalization dynamics on some Cayley type tree
was started in \cite{ott} with recent 
extensions done in \cite{ketzmerick,nechaev,agam}.
However, a detailed understanding of the intriguing
features of Poincar\'e recurrences in the
Chirikov standard map and other similar maps
is still missing.

In this work we use a generalized Ulam method
developed in \cite{frahmulam,qwlib} and combine it
with a new survival Monte Carlo method
trying to reach larger time scales and to obtain a better understanding of
statistics of Poincar\'e recurrences in the Chirikov standard
map and the separatrix map. 

The paper is composed as follows: in Section 2 we construct the Ulam
matrix based on the generalized Ulam method and study 
the properties of its spectrum, eigenstates and corresponding time 
evolution for the case of the Chirikov standard map. 
The survival Monte Carlo method is introduced in Section 3
and the properties of the Poincar\'e recurrences are studied
with its help comparing results with the Ulam method. 
In Section 4 we apply the above methods to the separatrix map and 
in Section 5 the localization properties of the eigenstates of the 
Ulam matrix are analyzed. The discussion of the results 
is presented in Section 6.

\section{Generalized Ulam method with absorption}

The Ulam method was proposed in 1960 \cite{ulam}.
In the original version of this method a 2D phase space is divided in
$N_d=M \times M$ cells and $n_c$ trajectories
are propagated on one map iteration from each cell $j$.
Then the matrix $S_{ij}$ is defined by the relation
$S_{ij}= n_{ij}/n_c$ where $n_{ij}$ is the number of trajectories
arriving from a cell $j$ to a cell $i$.
By construction we have $\sum_i S_{ij}=1$ and hence the
matrix $S_{ij}$ belongs to the class of the Perron-Frobenius
operators (see e.g. \cite{mbrin}). This Ulam matrix can be considered
as a discrete Ulam approximate of the Perron-Frobenius
operator (UPFO) of the continuous dynamics. 

\begin{figure}[h]
\begin{center}
\includegraphics[width=0.48\textwidth]{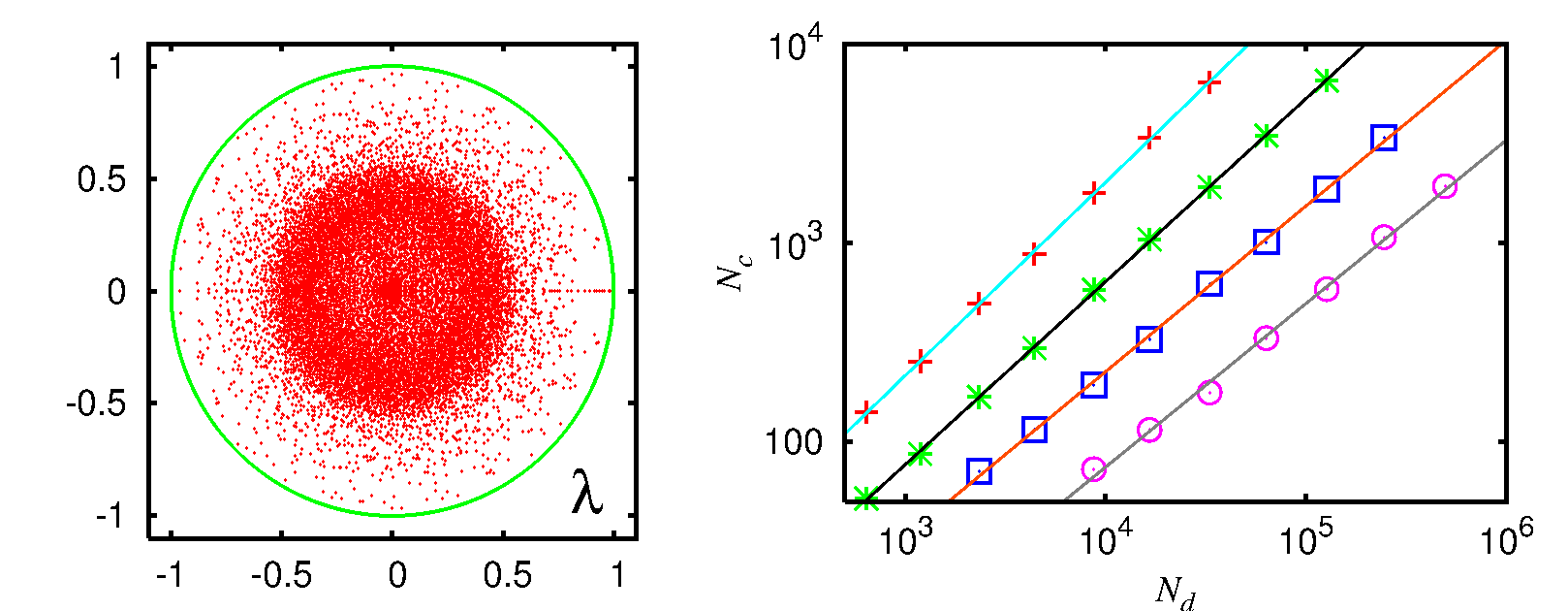}
\end{center}
\caption{\label{fig1} 
(Color online) {\it The left panel} shows the eigenvalue spectrum $\lambda_j$ 
for the projected case of the UPFO of map (\ref{eq_stmap}) at $K=K_g$ 
in the complex plane for $M=280$ and $N_d=16609$ 
by red/gray dots 
(projected matrix dimension $N_p=15457$). 
The green/gray curve represents the circle $|\lambda|=1$.
{\it The right panel} shows the number $N_c$ of eigenvalues, with 
modulus larger than $\lambda_c$, versus $N_d$ in a double logarithmic 
representation for $\lambda_c=0.5$ (crosses), $\lambda_c=0.66$ (stars), 
$\lambda_c=0.8$ (open squares) and $\lambda_c=0.9$ (open circles). The 
straight lines correspond to the power law fits $N_c\sim\,N_d^\nu$ with 
exponents 
$\nu=0.971 \pm 0.006$ ($\lambda_c=0.5$), 
$\nu=0.919 \pm 0.005$ ($\lambda_c=0.66$), 
$\nu=0.832 \pm 0.010$ ($\lambda_c=0.8$) and 
$b=0.821 \pm 0.021$ ($\lambda_c=0.9$). The fits are done for the 
data with $N_c>50$, $M>35$ and 
$M\le 400$ ($\lambda_c=0.5$), 
$M\le 800$ ($\lambda_c=0.66$), 
$M\le 1120$ ($\lambda_c=0.8$), 
$M\le 1600$ ($\lambda_c=0.9$),
 since the Arnoldi method provides only 
a partial spectrum of the eigenvalues with 
largest modulus for large values of 
$M$. 
}
\end{figure}

According to the Ulam conjecture \cite{ulam}
the UPFO converges to the continuous limit
at large $M$. Indeed, this conjecture was proven
for 1D homogeneously chaotic
maps \cite{li}. Various properties of the UPFO
for 1D and 2D maps are analyzed in 
\cite{tel,kaufmann},\cite{froyland2007,froylandphysd}. 
Recent studies 
\cite{zhirov,ermann} demonstrated similarities 
between the UPFO, the corresponding  to them Ulam networks 
and the properties of the Google matrix of the world wide web
networks. It was shown that in maps with absorption
or dissipation the spectrum of the UPFO is
characterized by the fractal Weyl law \cite{ermann1}.

\begin{figure}[h!]
\begin{center}
\includegraphics[width=0.48\textwidth]{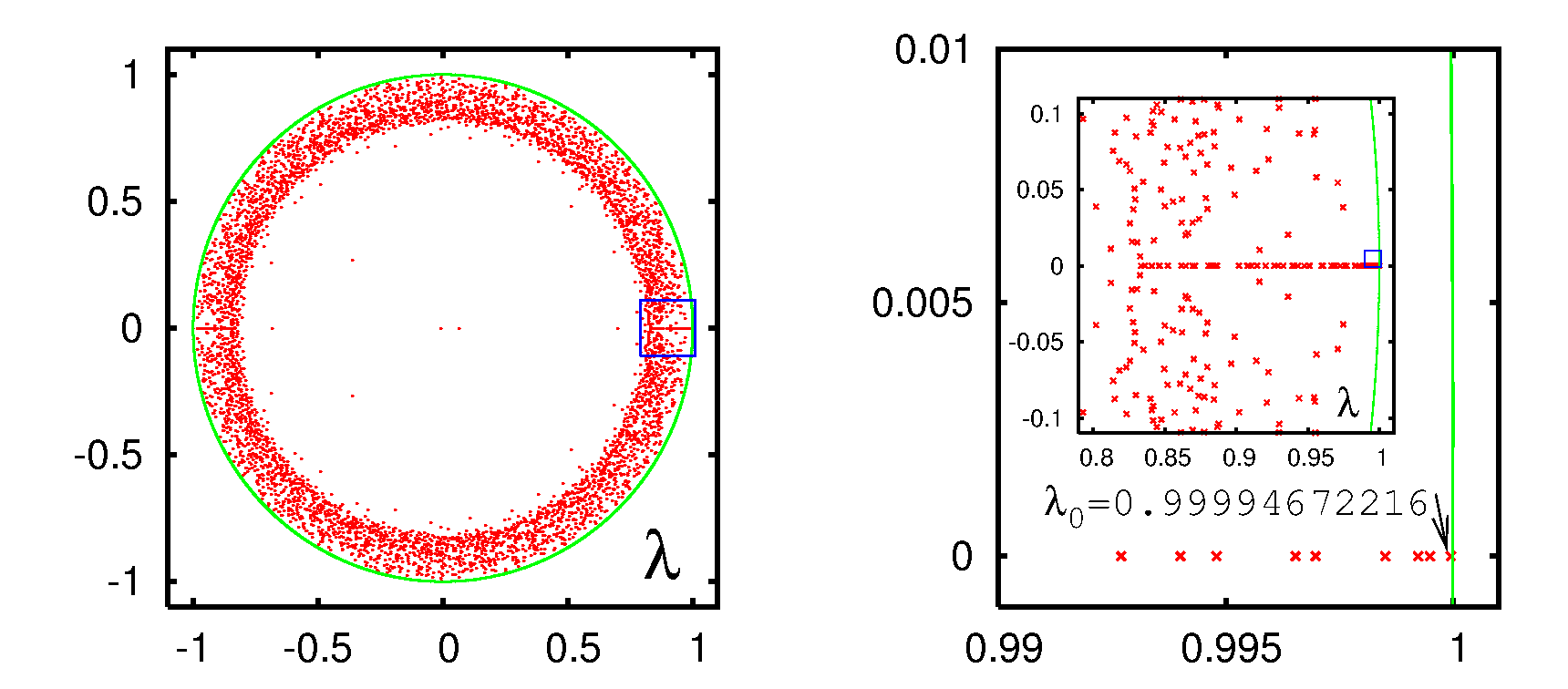}
\end{center}
\caption{\label{fig2} (Color online) Partial spectrum $\lambda_j$ 
for the projected case of the UPFO of the map (\ref{eq_stmap}) at $K=K_g$ 
for $M=1600$. 
{\it The left panel} shows all eigenvalues obtained by 
the Arnoldi method with $n_{\rm A}=5000$. 
The insert of {\it the right panel} shows the blue/black square of the first 
zoomed range of the {\it left panel};
 the blue/black square here is the 
second zoomed range shown in the main figure of {\it the right panel}. 
The eigenvalue 
with largest modulus $\lambda_0=0.99994672216$ is indicated 
by an arrow. The green/gray 
curve represents in all cases the circle $|\lambda|=1$. 
}
\end{figure}

The coarse-grained cell structure of the original Ulam method corresponds 
to an effective noise and in case of a divided phase space the noise 
induces an artificial diffusion between chaotic and regular regions. 
In \cite{frahmulam} this problem was solved by replacing the random 
initial points by a very long chaotic trajectory and 
the transitions between cells are accumulated along the chaotic trajectory
that keeps the invariant curves and stable islands
even in presence of the effective noise. Furthermore, the matrix size is also 
reduced since only cells which are visited at least once by the trajectory 
are kept. Here we use this approach for the 
analysis of the Poincar\'e recurrences keeping the same
notations as in \cite{frahmulam}. In particular, as in 
\cite{frahmulam}, we exploit the parity symmetry 
$x\to 1-x$ and $y\to 1-y$ allowing to limit the effective phase space 
to $0\le x\le 1$, $0\le y\le 0.5$ and therefore reducing the number of 
cells at a given cell size by a factor of two. In $x$ direction we use 
therefore $M$ cells and in $y$ direction $M/2$ cells with 
$M\in\{25,\,35,\,\ldots,\,1120,\,1600\}$
and the intermediate values are multiples of $25$ or $35$ by powers of 2.

To study the Poincar\'e recurrences withing the Ulam method
we introduce absorption of all trajectories
with $y<y_{cut}=0.05$. Thus we generate 
the matrix $S$ using one trajectory iterated by the map up to 
the iteration time $t=10^{12}$
(as in \cite{frahmulam}; this corresponds to
the closed system without absorption and we call this
the symplectic case). After that the matrix size $N_d$ is simply reduced
only to those cells with $y \ge y_{cut}$ that gives the projected
matrix dimension $N_p$ and matrix $S_p$. 
The matrix size of this projected case
is smaller approximately by $7 \%$. 
We find, for $M \leq 1600$, an approximate dependence
$N_d \approx 0.39 M^2/2$ and
$N_p \approx 0.36 M^2/2$.
This corresponds to the usual estimate of the chaos
measure being around $39 \%$ in agreement with the
results of Chirikov \cite{chirikov1979}. 
For the symplectic case 
we have the maximal eigenvalue $\lambda =1$ while in the projected case
with absorption we are getting $|\lambda|<1$.

The spectrum $\lambda_j$ of the projected case with matrix $S_p$ is shown
in Fig.~\ref{fig1}. The spectrum is obtained 
by the direct diagonalization of the matrix $S_p$
that can be done numerically up to $M=280$.
It can be compared with the corresponding spectrum
of the symplectic system shown in Fig.~2 of \cite{frahmulam}.
The global spectrum structure of $S$ for the symplectic case is similar to
the projected case. Indeed, the absorption is relatively weak
and does not affect  the global properties of motion.
However, with absorption the measure is not conserved
and the remaining non-escaping set forms a fractal
set with the fractal dimension $d<2$ (see e.g. \cite{ermann1,dls}).

In the case of Ulam networks on fractal chaotic repellers
the spectrum of UPFO $S_p$ is characterized by the fractal Weyl law
with the number of states $N_c$ in the ring
$\lambda_c < |\lambda| \leq 1$ growing with the matrix size
$N_d$ as $N_c \propto N_d^\nu$ (here for simplicity we use the size $N_d$
of the symplectic case, for the projected case we have 
simply to change $N_p \approx 0.93 N_d$). It can be argued that
the fractal dimension $d_0$ of the invariant repeller set 
determines the exponent $\nu=d_0/2$ \cite{ermann1}.
Examples of dependencies $N_c$ vs $N_d$ are given in Fig.~\ref{fig1}
for various values of $\lambda_c$.
Definitely we have $\nu<1$ but there is an evident dependence on
$\lambda_c$ with a decreasing value of $\nu$
at $\lambda_c \rightarrow 1$. We attribute this to the fact
that at $\lambda_c \rightarrow 1$ we are dealing with long sticking
trajectories whose measure decreases with time.
\begin{figure}[h]
\begin{center}
\includegraphics[width=0.48\textwidth]{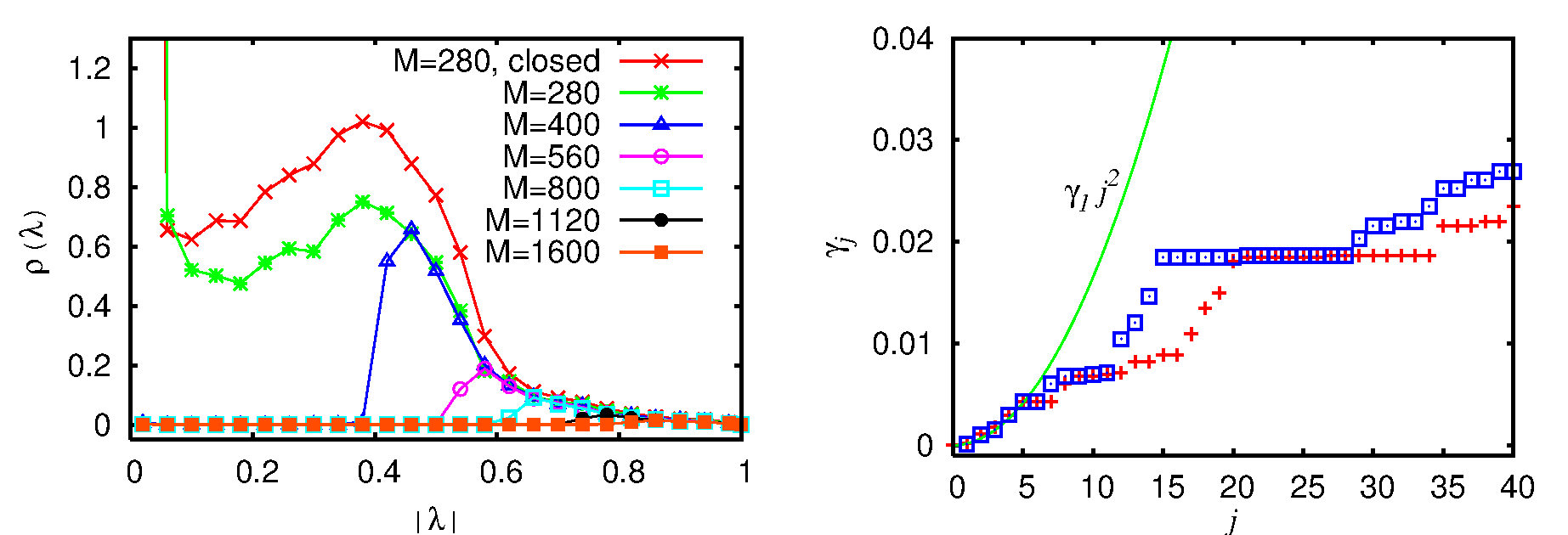}
\end{center}
\caption{\label{fig3} (Color online) 
{\it The left panel} shows the density $\rho(\lambda)$
of eigenvalues, being normalized by $\int \rho(\lambda)\,d^2\lambda = 1$, 
of the UPFO for the map (\ref{eq_stmap})
at $K=K_g$ in the complex plane as a function of the modulus 
$|\lambda|$ for $M=280$ for the symplectic case (upper curve, crosses) and 
the projected case (lower green curve, stars). 
The other curves are partial (non-normalized) densities for the projected 
case and the values $M=400,\,560,\,800,\,1120,\,1600$ and the number of used 
eigenvalues (obtained by the Arnoldi method) is 
$n_A=12000,\,8000,\,8000,\,6000,\,5000$ respectively.
{\it The right panel} shows the 
decay rates $\gamma_j=-2\ln(|\lambda_j|)$ versus level number $j$ for the 
UPFO eigenvalues $\lambda_j$, with $M=1600$ and $N_d=494964$. 
The red/gray crosses correspond 
to the  UPFO of symplectic case and the blue/black squares correspond 
to the projected case (data points for this case are shifted to 
one position to the right). 
The green curve corresponds to the 
quadratic dispersion law 
$\gamma_j\approx \gamma_1\,j^2$ which is approximately valid for the 
diffusion modes with $0\le j\le 5$ and where $\gamma_1$ is taken from 
the UPFO of the symplectic case. 
}
\end{figure}

Here we should point out that the data for $M\ge 400$ corresponding 
to $N_d >30000$
are obtained from the Arnoldi method \cite{arnoldibook}
which allows to find the eigenvalues for matrix sizes up to $N_d \sim 10^6$.
However, only a finite number of eigenvalues with 
largest $|\lambda|$ can be determined numerically using 
$n_A=12000,\,8000,\,8000,\,6000,\,5000$ (for 
$M=400,\,560,\,800,\,1120,\,1600$ respectively and 
with $n_A$ being the used Arnoldi dimension).
A more detailed description of the Arnoldi method for the 
UPFO is given in \cite{frahmulam}. An example of the spectrum 
$\lambda$ obtained with the Arnoldi method at 
the largest value of $M=1600$ is shown in Fig.~\ref{fig2}.
Here $N_d=494964$ and $N_p=458891$.
We find that the maximal eigenvalue for the
projected case is $\lambda_0=0.99994672216$
corresponding to a slow escape rate at large times. As in \cite{frahmulam} 
for the symplectic case without absorption, we obtain also for the case with 
absorption two type of eigenmodes: ``diffusion modes''
with real eigenvalues close to 1 and whose eigenvectors are rather extended in 
phase space (with some decay for cells close to the absorption border) 
and ``resonant modes'' with complex or real negative eigenvalues and 
which are quite well localized around a chain of 
stable islands close to an invariant curve. It turns out that 
many of the resonant modes (those ``far'' away from the absorption border), 
coincide numerically very well with corresponding resonant modes for 
the case without absorption already found in \cite{frahmulam}.

The dependence of the density of eigenvalues $\rho(|\lambda|)$ 
on $|\lambda|$ is shown in Fig.~\ref{fig3}.  
We see the proximity between the symplectic and projected cases
not only in density $\rho$ but also in a slow relaxation of the 
diffusion modes with relaxation rates $\gamma_j \approx \gamma_1 j^2$
($\gamma_j=-2 \ln |\lambda_j|$) provided we identify $\gamma_{j+1}$ 
of the symplectic case with $\gamma_j$ of the projected case because 
$\gamma_0$ of the symplectic case is simply zero and 
the relaxation rate $\gamma_1$ to the ergodic state of the symplectic 
case corresponds roughly to the exponential long time escape rate $\gamma_0$ 
of the projected case. 
The proximity of the two cases is also well
seen in the dependence of  integrated density of states
$\rho_\Sigma(\gamma)= j/N_d$ on $\gamma_j$ shown in Fig.~\ref{fig4}
(here $j$ is a number of eigenvalues with
$\gamma \leq \gamma_j$).
In both cases we have the algebraic dependence
$\rho_\Sigma(\gamma) \propto \gamma^\beta$ with 
$\beta \approx 1.5$. In \cite{frahmulam} it was argued that
this exponent is the same as for the exponent
of decay of Poincar\'e recurrences $P(t)$.
These data show that an introduction of small absorption
at $y<y_{cut}$ does not produce significant 
modification for trajectories trapped for long times
in a vicinity of the critical golden curve or other
secondary islands located far away from the absorption band. 

\begin{figure}[h]
\begin{center}
\includegraphics[width=0.48\textwidth]{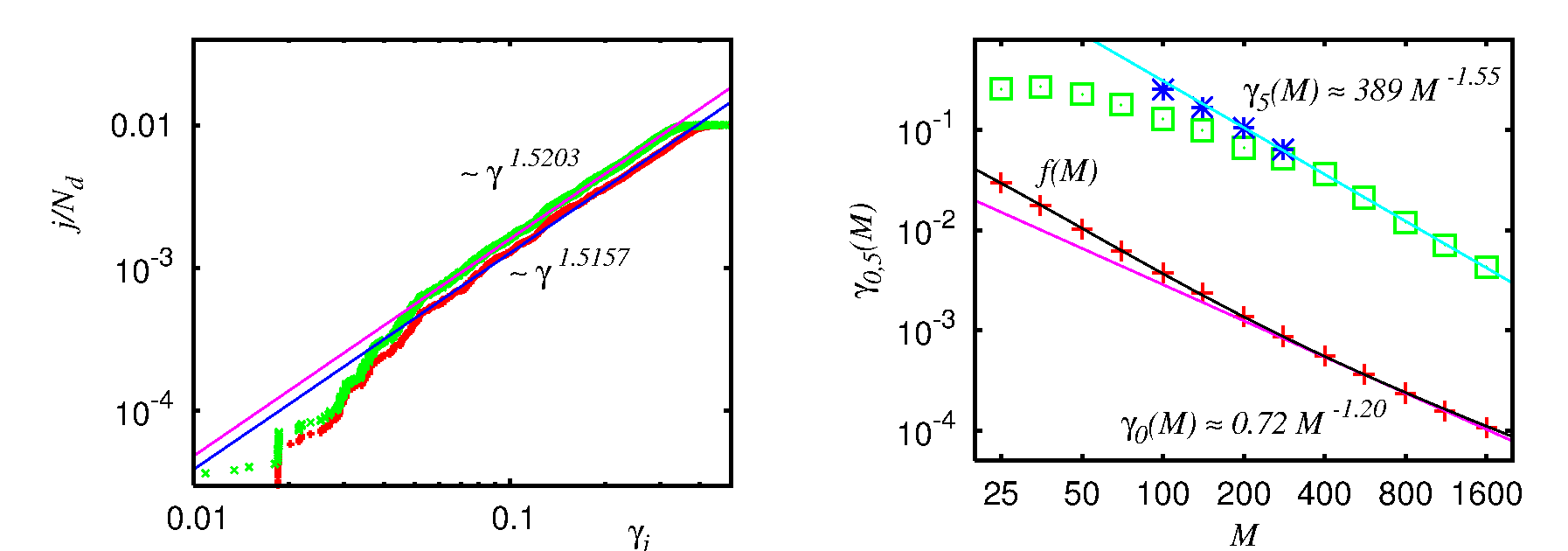}
\end{center}
\caption{\label{fig4} (Color online)
{\it The left panel} shows the rescaled level number $j/N_d$ versus the decay 
rate $\gamma_j$, in a double logarithmic scale,
for the map (\ref{eq_stmap}) at $K_g$ with $M=1600$ and $N_d=494964$. 
Red/lower data points correspond to the UPFO projected case 
and green/upper data points correspond to the UPFO symplectic case.
The two straight lines correspond to the power law fits
$j/N_d\approx 0.052745\,\gamma^{1.5203}$ (symplectic case) and
$j/N_d\approx 0.041570\,\gamma^{1.5157}$ (projected case) for
the data in the range $0.04\le\gamma\le 0.3$.
The statistical error bound of the exponents obtained from the fits
is close to $0.1\%$ in both cases. 
{\it The right panel} shows the decay rates $\gamma_j(M)$ for $j=0$ 
(red crosses), $j=5$ (green open squares) of the UPFO projected case 
in a double logarithmic scale. 
The lower/pink straight line corresponds to the power law fit 
$\gamma_0(M)\approx 0.72\,M^{-1.20}$ and the upper/light blue straight 
line to the fit $\gamma_5(M)\approx 389\,M^{-1.55}$ (both fits obtained 
for the range $400\le M\le 1600$). The black/curved line 
corresponds to the other fit 
$\gamma_0(M)=f(M)=\frac{D}{M}\,\frac{1+C/M}{1+B/M}$ 
with $D=0.162$, $C=165$ and $B=17.0$ (fit obtained for the 
range $25\le M\le 1600$). We mention that 
$\gamma_5$ corresponds for $M\ge 400$ to a resonant mode 
whose eigenvector is strongly localized close to the three stable islands 
of the resonance $1/3$. 
However, for $M\le 280$ $\gamma_5$ corresponds to a different mode 
and the resonant mode at $1/3$ is associated to $\gamma_7$ ($M=280$), 
$\gamma_{13}$ ($M=200$), $\gamma_{17}$ ($M=140$) and $\gamma_{23}$ ($M=100$) 
which are shown as four additional data points (blue stars).
}
\end{figure}

The lowest eigenvalues such like $\gamma_0$ and $\gamma_5$
decrease algebraically with the increase of $M$
as it is shown in right panel of Fig.~\ref{fig4}.
In the fit range $400\le M\le 1600$ we have a power law 
$\gamma_0(M)\approx 0.72\,M^{-1.20}$ but taking into account the curvature 
for the interval $25\le M\le 1600$ the modified fit 
$\gamma_0(M)=\frac{D}{M}\,\frac{1+C/M}{1+B/M}$ 
with $D=0.162$, $C=165$ and $B=17.0$ seems to indicate a behavior 
$\gamma_0(M)\propto M^{-1}$ in the limit $M\to\infty$. 
This behavior is similar to the one found in \cite{frahmulam}
for $\gamma_1$ in the symplectic case (where $\gamma_0$ is simply 0). 
On the other hand the resonant mode $\gamma_5$ obeys the power law 
$\gamma_5(M)\approx 389\,M^{-1.55}$ which is valid for the interval 
$100\le M\le 1600$ if we use for the smaller values of $M$ not $\gamma_5$ 
but the resonant mode localized to the same chain of resonant islands 
which may have a different eigenvalue index (see Fig.~\ref{fig4} for details). 
The comparison of these decays indicate that eventually at very large values 
of $M$, far outside the range numerically accessible by the Arnoldi method, 
the resonant modes become dominant over the diffusion modes. 
The limit $\gamma \rightarrow 0$ for $M\to\infty$ 
is related to long sticking trajectories near 
critical invariant curves which restrict the chaos component and whose 
phase space structure can be better resolved with decreasing cell 
size $1/M$. 
As in  \cite{frahmulam}
we argue that these lowest modes are affected by the effective
noise present in the Ulam method. Due to that we do not have a clear
explanation for this algebraic decay. However, the fact that 
$\gamma_j$ (at fixed value of $j$) vanishes with increasing $M$ 
indicates that the limit $t_{\rm exp}$ in time, when the statistics of 
Poincar\'e recurrences $P(t)$ 
obtained from the UPFO becomes exponential, increases 
as well according to $t_{\rm exp}\propto \gamma_0^{-1}$ and therefore we 
expect to recover the power law decay of $P(t)$ for $M\to\infty$ (see below in 
Section 3). 

\begin{figure}[h]
\begin{center}
\includegraphics[width=0.48\textwidth]{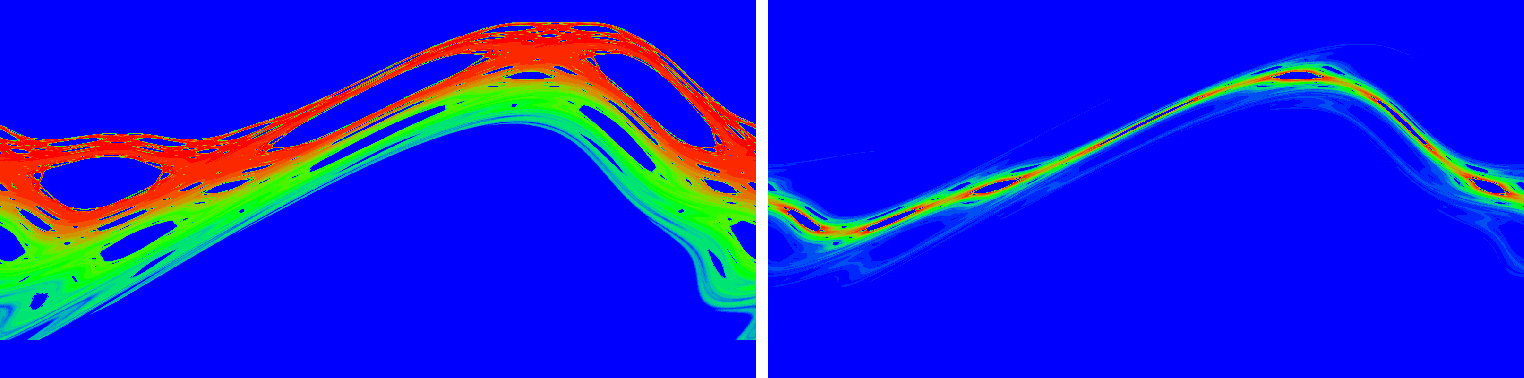}
\end{center}
\caption{\label{fig5} (Color online)
Density plot of the modulus of the eigenvector components 
$|\psi|$ of the UPFO 
projected case of  map (\ref{eq_stmap})
at $K_g$ with $M=1600$ for the two modes 
with eigenvalues $\lambda_0=0.99994672$ 
(left panel) and 
$\lambda_{29}=-0.22008951+i\,0.96448508\approx |\lambda_{29}|\,e^{i\,2\pi(2/7)}$
(right panel). The density is shown by color with red/gray for maximum 
and blue/black for zero.
}
\end{figure}

\begin{figure}[h]
\begin{center}
\includegraphics[width=0.48\textwidth]{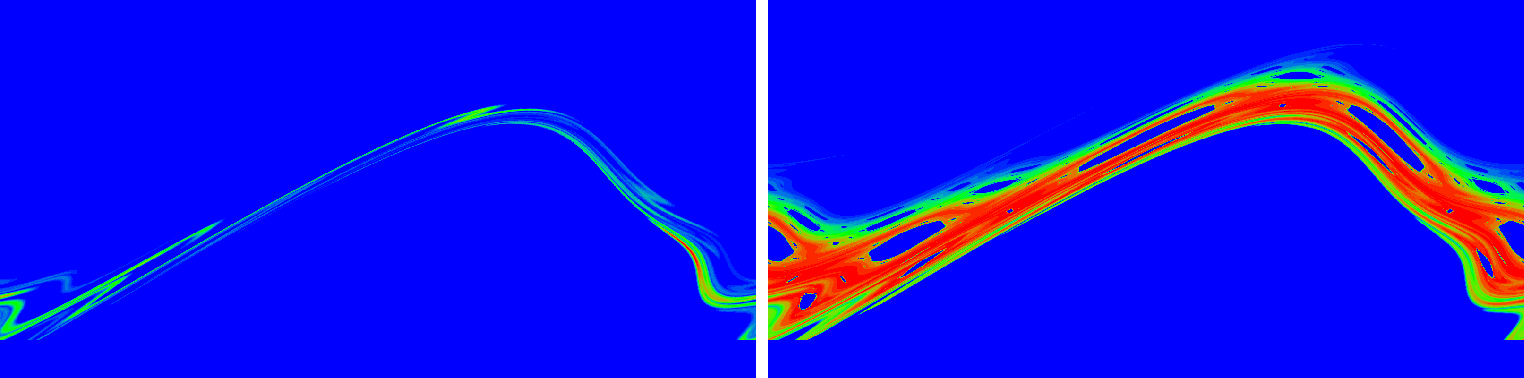}
\end{center}
\caption{\label{fig6} (Color online)
Time dependent probability density calculated by 
$\psi(t)={(S_p)}^t\,\psi(0)/\parallel {(S_p)}^t\,\psi(0)\parallel_1$
 where $S_p$ is the UPFO
for the projected case for $M=1600$, $\psi(0)$ an initial vector with 
$\psi_l(0)=\delta_{l,\ell_0}$ 
and $\ell_0$ being the index of the cell at $x_0=y_0=0.0625$ and 
$\parallel \cdots \parallel_1$ is the 
$1$-norm defined by $\parallel \psi \parallel_1=\sum_l |\psi_l|$. 
The densities are shown for $t=40$ (left panel) and 
$t=400$ (right panel). In the limit $t\to\infty$ the vector $\psi(t)$ 
converges to the eigenvector of maximal eigenvalue $\lambda_0$ shown 
in the left panel of Fig.~\ref{fig5}. The full convergence 
is achieved for $t\ge 40000$ so that  for these times 
the density plot of $\psi(t)$ remains unchanged at the given color-resolution.
}
\end{figure}

With the help of the Arnoldi method we find certain eigenstates
corresponding to eigenvalues of the matrix $S_p$ and satisfying the equation
\begin{equation}
\label{eq_eigen}
\sum_{i=0}^{N_p-1} (S_{p})_{mi} \psi_j(i) = \lambda_j\psi_j(m)  \; .  
\end{equation}
Examples of two eigenmodes $|\psi_0|$ and 
$|\psi_{29}|$ are shown in Fig.~\ref{fig5}.
The state $|\psi_0|$ corresponds to the first diffusive mode
mainly located in a vicinity of the critical golden curve
while $|\psi_{29}|$ corresponds to the mode located near a resonant 
chain with rotation number $r=2/7$.

It is also interesting to follow how the probability 
initially placed in one cell $\ell_0$ evolves
with time. Of course, the total probability 
starts to decay due to absorption but by renormalizing the 
total probability back to unity after each map iteration
we obtain its evolution in phase space.
At large times we have convergence to the state $\psi_0$ with maximal
$\lambda_0$ but at intermediate times we see the regions of phase
space which contribute to long time sticking and long
Poincar\'e recurrences. Two snapshots are shown in Fig.~\ref{fig6}.
The videos of such an evolution for the maps (\ref{eq_stmap})
and  (\ref{eq_separatrixmap}) are available at \cite{qwlib}.

\section{Poincar\'e recurrences \\ by survival Monte Carlo method} 

The numerical computation of the Poincar\'e recurrences
counting the number of crossing of a given line (e.g. $y=0$)
in the phase space is known to be a very stable numerical method
since the integrated probability of recurrences on a line
at times larger than $t$ is positively defined
(see e.g. \cite{kiev,kievb},\cite{chirikov1999,ketzmerick}).
However, at large times the direct numerical computation
becomes time consuming. 

With the aim to reach larger
times we  present here a new method to calculate 
the statistics of Poincar\'e recurrences of map such as the Chirikov standard 
map (\ref{eq_stmap}). We will call this method 
the {\em Survival Monte Carlo method} (SMCM). 
The idea of this method is to chose a certain, quite large number 
$N_i\gg 1$, of initial conditions randomly chosen in some small cell close 
to an unstable fix point and 
to calculate in parallel the time evolution of these trajectories. 
At the initial time $t=0$ we put the Poincar\'e return 
probability to $P(0)=1$ and the number of trajectories to $N(0)=N_i$. 
At each time $t_k$, when a given trajectory escapes 
in the absorption region $y<y_{cut}=0.05$ 
of the phase space,  we put 
$P(t_{k}+1)=P(t_k)\,(N(t_k)-1)/N(t_k)$ and
 $N(t_{k}+1)=N(t_k)-1$,
otherwise we simply keep
$P(t_{k}+1)=P(t_k)$ and $N(t_{k}+1)=N(t_k)$. 
When the number of remaining trajectories 
$N(t_k)$ drops below a certain threshold value $N_f$ 
(typically chosen such that 
$N_i\gg N_f\gg 1$) we {\em reinject} a new trajectory close to one of the 
other remaining trajectories with a small random deviation: 
$x_{\rm new}(t)\in [x_i(t)-\varepsilon/2,\,x_i(t)+\varepsilon/2]$ 
and 
$y_{\rm new}(t)\in [y_i(t)-\varepsilon/2,\,y_i(t)+\varepsilon/2]$. 
The main idea is to keep a typical statistics of trajectories at a given 
time $t$ and to concentrate the computational effort on the very long and rare 
trajectories without wasting resources on the more probable 
trajectories with short times of Poincar\'e recurrences.

\begin{figure}[h]
\begin{center}
\includegraphics[width=0.48\textwidth]{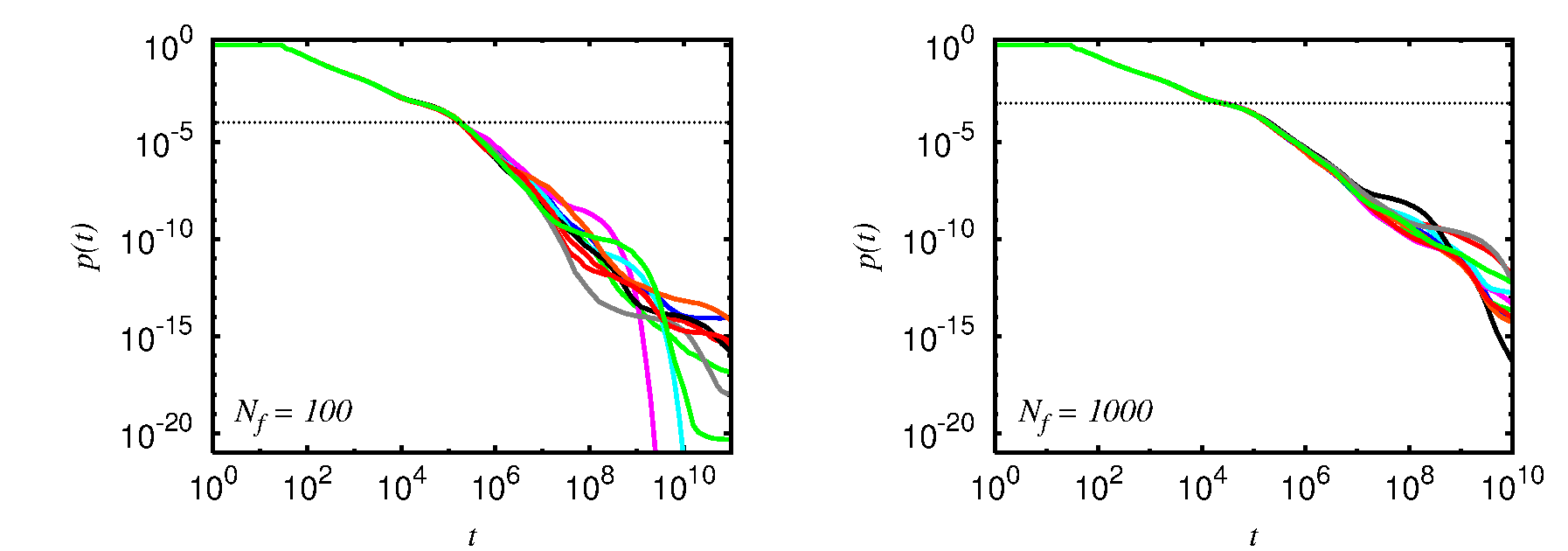}
\end{center}
\caption{\label{fig7} (Color online)
Statistics of Poincar\'e recurrences $P(t)$ of the map (\ref{eq_stmap})
calculated by the SMCM as survival probability after times larger than $t$
(data are shown in double logarithmic scale). The 
number of initial trajectories is $N_i=10^6$ and the number of final 
trajectories is $N_f=100$ (left panel) or $N_f=1000$ (right panel). 
The initial positions are randomly chosen in a cell of size 
$(1600)^{-1}\times (1600)^{-1}$ at the position $x_0=y_0=0.0625$,
here the small random deviation for reinjected trajectories is
$\sim\varepsilon=10^{-14}$.  
In both panels the results for $P(t)$ are shown for 10 realizations with 
different random seeds. 
The horizontal dotted line indicates the limit probability $N_f/N_i=10^{-4}$ 
(left panel) or $N_f/N_i=10^{-3}$ (right panel) below which the reinjection of 
trajectories is applied.
The two realizations in the left panel which drop below the shown range (of 
$P(t)\ge 10^{-21}$) ``saturate'' eventually at the values 
$P(10^{11})\approx 2\times 10^{-36}$ or $P(10^{11})\approx 10^{-35}$.
} 
\end{figure}

In this method the proper choice of $\varepsilon$ is important. On 
one hand $\varepsilon$ should not be too small in order to avoid too strong 
correlations between the trajectories and on the other hand it should be 
very small in order to avoid an uncontrolled too strong diffusion into 
regions too close to stable islands where the trajectories may be trapped 
stronger and longer as they should be without the random deviations. 
Fortunately in the chaotic region even a modest Lyapunov exponent ensures 
exponential separation of trajectories and choosing a very small value of 
$\varepsilon$ one may hope to reduce the correlation between the injected 
trajectory and its reference trajectory after a modest number of iterations. 
Furthermore at longer times the average time between the escape of two 
trajectories becomes very large that helps to reduce these correlations. 

\begin{figure}[h]
\begin{center}
\includegraphics[width=0.48\textwidth]{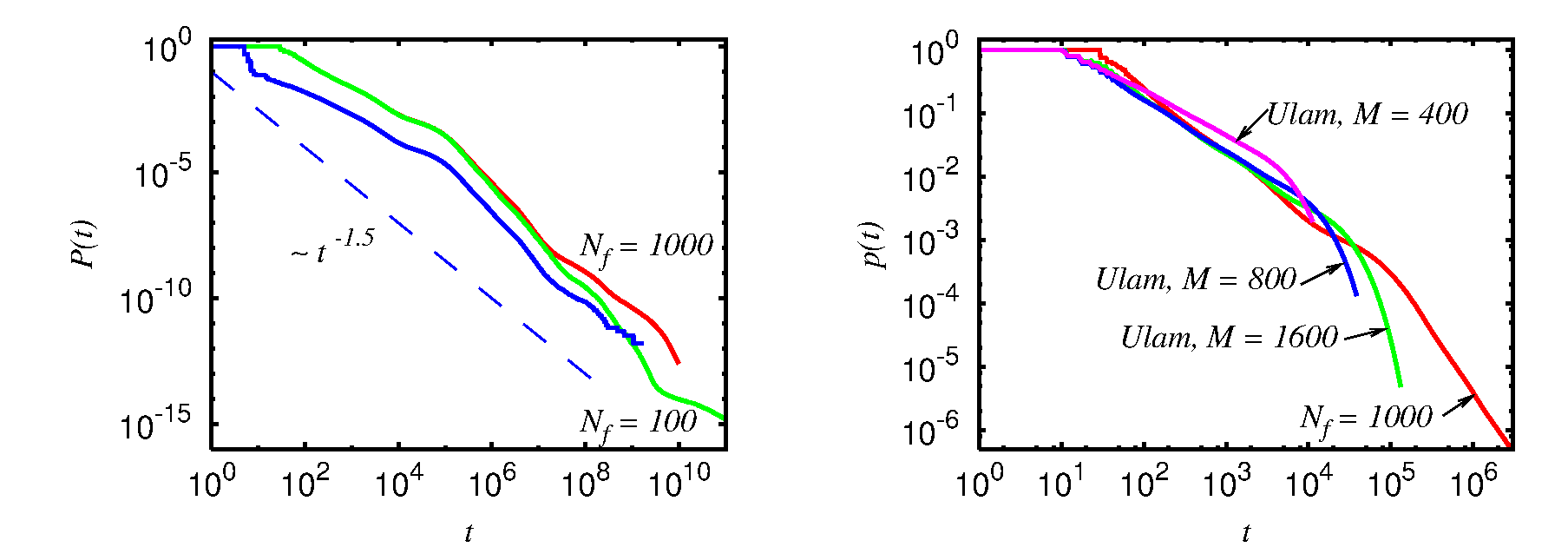}
\end{center}
\caption{\label{fig8} (Color online)
{\it The left panel} shows the average over  10 random realizations 
of the statistics of Poincar\'e recurrences $P(t)$ of the map 
(\ref{eq_stmap}), obtained by the 
SMCM for survival probability and shown in Fig.~\ref{fig7}. 
The upper/red curve, for $t\le 10^{10}$, corresponds to $N_f=1000$. 
The next lower/green curve, for $t\le 10^{11}$, corresponds 
to $N_f=100$. The lowest blue curve, for $t\le 1.7\times 10^9$, 
corresponds to the data of Refs.~\cite{chirikov1999b,ketzmerickcom} 
obtained by a direct computation of the 
statistics of Poincar\'e recurrences.
The dashed straight line indicates a power law behavior $P \propto t^{-1.5}$. 
{\it The right panel} compares the statistics of Poincar\'e recurrences 
$P(t)$ obtained by the SMCM for $N_f=1000$ to $P(t)$ obtained by the 
Ulam method for $M=400,\,800,\,1600$. At  large times 
$t>t_{\rm exp}\sim 10^4-10^5$ the curves obtained by the Ulam 
method show an exponential behavior $P(t)\sim \lambda_0^t$ 
determined by the largest eigenvalue of the UPFO for the projected case.
}
\end{figure}

We have chosen the parameters $\varepsilon=10^{-14}$, $N_i=10^6$ and the 
two cases $N_f=100$ and $N_f=1000$. For $N_f=100$ we have been able 
to iterate up to times $10^{11}$ and for $N_f=1000$ up to times $10^{10}$. 
We mention that at the ``larger'' value $\varepsilon=10^{-10}$, we observe 
a significant effect of artificial saturation in $P(t)$ (i.~e. no escapes of 
trajectories) at longer times because the 
trajectories penetrate ``too strongly'' into the regions very close to the 
stable islands or the critical curve. When we choose 
$\varepsilon=10^{-14}$ this artificial saturation effect is strongly reduced. 
We  calculate in parallel different realizations of $P(t)$ 
with respect of the random variables (for the initial conditions, for 
the random deviations of the reinjected trajectories and for the random 
choice at which remaining trajectory the reinjection happens). 
The comparison of obtained data shows that the distribution
$P(t)$ is stable at small and large times. But at very large times
it turns out that the fluctuations become quite strong. 

Examples of the survival probability $P(t)$ obtained 
for 10 different realizations with $N_f=100$ (left panel) and
$N_f=1000$ (right panel) are shown in Fig.~\ref{fig7}.
Of course the fluctuations appear for $N_f=100$ 
at shorter times ($t\sim 10^5-10^6$)
as compared to $N_f=1000$ ($t\sim 10^6-10^7$). 

\begin{figure}[h]
\begin{center}
\includegraphics[width=0.48\textwidth]{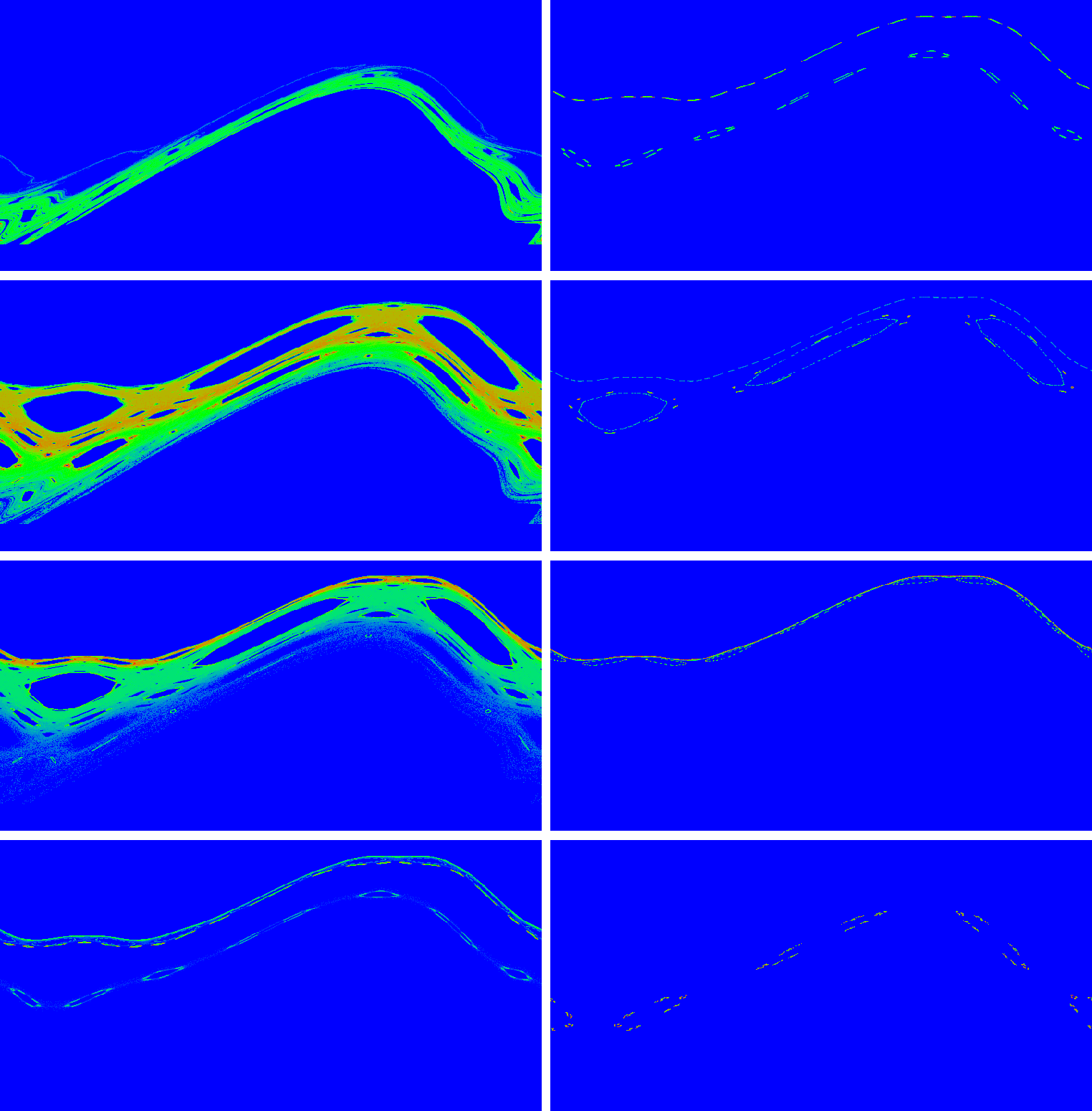}
\end{center}
\caption{\label{fig9} (Color online)
Density plots of the trajectories of the SMCM (with $N_f=1000$) 
for the map (\ref{eq_stmap}) 
for various times $t$ and random realizations. All 
density plots are obtained from a histogram of $10^7$ data points 
and using a resolution of $800\times 400$ cells for the phase space 
$0\le x< 1$ and $0\le y<0.5$. The data points are obtained 
by iterating $N(t)$ trajectories (with $N(t)=P(t)\,N_i$ for 
$P(t)\ge 10^{-3}$ and $N(t)=N_f$ for $P(t)<10^{-3}$) from $t$ to 
$t+\Delta t$ with $\Delta t=10^7/N(t)$. The left four panels and the upper 
right panel correspond to one particular random realization at 
$t=10^2,\,10^4,\,10^6,\,10^8,\,10^{10}$ and the three lower right 
panels correspond to three other random realizations at $t=10^{10}$. 
For short times $t<10^{5}$ there is no significant difference between the 
density plots for different random realizations at a given time. 
More detailed density plots for  intermediate times and higher resolution 
figures are available at \cite{qwlib}.
}
\end{figure}

We note that the SMCM allows us to determine the survival probability
$P(t)$. Its comparison with the statistics of Poincar\'e
recurrences computed by the usual method \cite{kiev,chirikov1999},
\cite{chirikov1999b,ketzmerickcom} is shown in Fig.~\ref{fig8}.
We see that both methods give the same behavior $P(t)$
with a small shift in time related to different initial conditions.
The equivalence of both methods is rather clear:
in both methods the probability is determined by
long sticking trajectories; both methods consider the recurrences to
the lines $y=0$ or $y=0.05$ which are close to each other.

The decay of $P(t)$ averaged over 10 random realizations
is shown in Fig.~\ref{fig8}.
In general we see that the SMCM allows to reach extremely long
times with $t=10^{11}$ for $N_f=100$
and $t=10^{10}$ for $N_f=1000$. 
For $N_f=100$ we see that the fluctuations start to be important
for $t>10^9$ while the case with $N_f=1000$ remains stable
up to $t=10^{10}$. This allows to obtain the behavior of $P(t)$
for times being about one order of magnitude larger
compared to previous numerical simulations.

For the case $N_f=1000$ in Fig.~\ref{fig8}
the algebraic fit of data
in the range  $10^6 \leq t \leq 10^{10}$ gives
the Poincar\'e exponent $\beta=1.587 \pm 0.009$.
For $N_f=100$ case we find
$\beta=1.710 \pm 0.017$ for the range $10^6 \leq t \leq 10^{11}$.
The formal statistical error is rather small in both
cases but it is clear that for $N_f=100$
we start to have an effect of strong fluctuations
due to long sticking around islands
and thus the reliable value of $\beta$ is given by the case with
$N_f=1000$.

The survival probability $P(t)$ can be also computed using the Ulam method
at various sizes of discrete cells determined by $M$.
The results obtained by the generalized Ulam method
and by the SMCM are shown in the right panel of Fig.~\ref{fig8}.
The comparison shows that both methods give the same results
but the SMCM is much more efficient allowing to follow the decay $P(t)$
up to significantly larger times since for the Ulam method we expect 
the decay $P(t)$ only to be accurate for $t<t_{\rm exp}\sim\gamma_0^{-1}$
because for $t>t_{\rm exp}$ it becomes exponential 
$P(t)\propto \lambda_0^t=\exp(-\gamma_0\,t/2)$. The data of 
Fig.~\ref{fig8} clearly shows that $t_{\rm exp}$ increases with $M$ 
in accordance with the decay of $\gamma_0$ obtained from Fig.~\ref{fig4}. 

Using the SMCM we can follow the evolution of the survival probability
as a function of time showing the density plot of long sticking trajectories.
Examples of such distributions are shown in Fig.~\ref{fig9}.
These Figs. show that at short times $t<100$ the trajectories are not
yet able to cross the cantori barriers and remain relatively far 
from the golden curve, at larger times $t=10^4, 10^6, 10^8$
the probability becomes concentrated close to the golden curve.
But at very larger times $t=10^{10}$ we find trajectories 
sticking in a vicinity of the golden curve or other secondary
resonances. Thus we see that at long time $P(t)$ has contributions
not only from the vicinity of the critical golden curve
but also from other secondary resonances. In this respect,
our conclusion confirms a similar one expressed in \cite{ketzmerickpre}
obtained from simulations on shorter time scales.

\section{Separatrix map with critical golden curve}

To show that the previous case of the Chirikov standard map
represents a generic situation we also study the UPFO 
of the projected case for the separatrix map \cite{chirikov1979}, defined by~:
\begin{equation}
\label{eq_separatrixmap}
{\bar y} = y + \sin (2\pi x) \; , \;\; 
{\bar x} = x + \frac{\Lambda}{2\pi}\ln(|{\bar y}|) \;\; ({\rm mod} \; 1) \;.
\end{equation}
This map can be locally  
approximated by the Chirikov standard map by linearizing the logarithm 
near a certain $y_0$ that leads after rescaling to the map (\ref{eq_stmap})
with an effective parameter 
$K_{\rm eff}=\Lambda/|y_0|$ \cite{chirikov1979}. 
As in \cite{frahmulam} we  study the map (\ref{eq_separatrixmap})
at $\Lambda_c=3.1819316$ with the critical 
golden curve at the rotation number
$r=r_g=(\sqrt{5}-1)/2  = 0.618..$. The construction of 
the matrix $S$ is described in \cite{frahmulam}, its size is
given by an approximate relation $N_d \approx 0.78 M^2/2$
for the phase space region
$0 < x \leq 1$, $0 \leq y \leq 4$ (symplectic case and 
using the symmetry: $x\to x+1/2\ ({\rm mod}\ 1)$, $y\to -y$). 
The absorption is done for $y < y_{cut}=0.4$ corresponding to 
10\% of the maximal possible value of $y$. 
Thus for the UPFO for the projected case we have
$N_p \approx 0.68 M^2/2$. In fact we have $2(N_d-N_p)/M^2=0.1$
since all part of the phase space is chaotic at
$0 <y <y_{cut}$ and all cells in this region were occupied by the 
Ulam method. Thus for 
$M=1600$ we have $N_d=997045$, $N_p=869045$.
\begin{figure}[h]
\begin{center}
\includegraphics[width=0.48\textwidth]{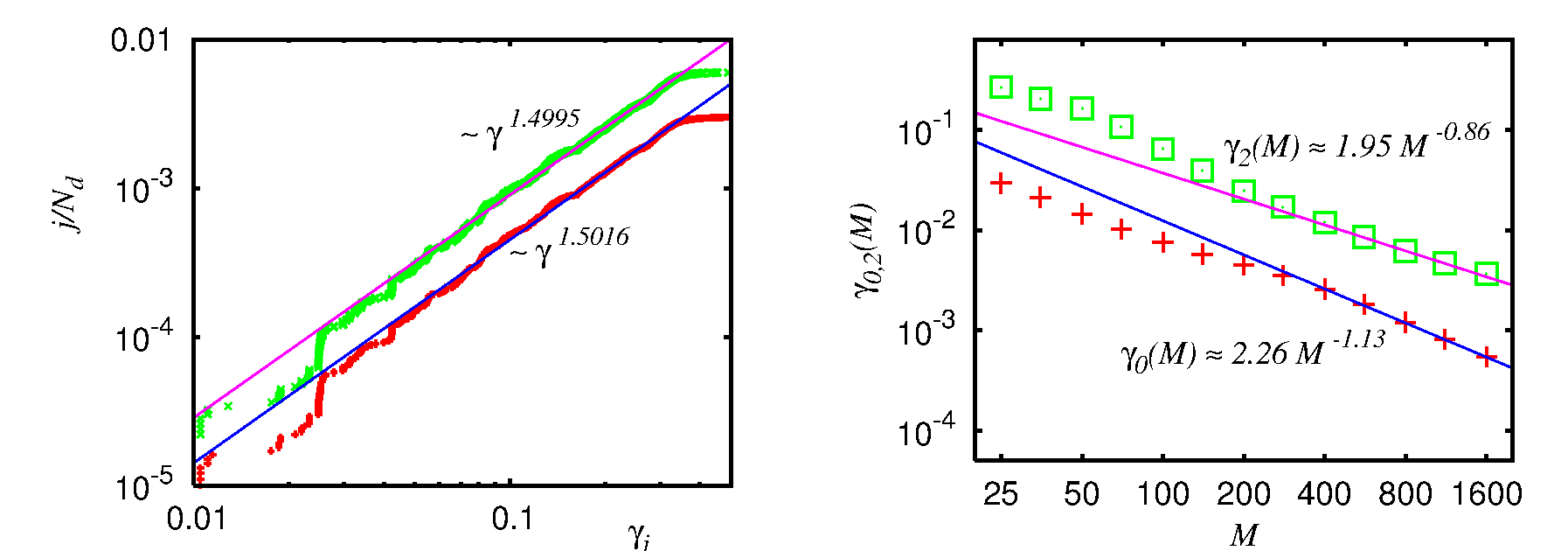}
\end{center}
\caption{\label{fig10} (Color online)
{\it The left panel} shows the rescaled level number $j/N_d$ versus the decay 
rate $\gamma_j$, in a double logarithmic scale,
for the separatrix map (\ref{eq_separatrixmap}) at $\Lambda_c$ with $M=1600$ 
and $N_d=997045$. 
Red/lower data points correspond to the UPFO for the projected case 
and green/upper data points correspond to the symplectic case. 
For the symplectic case the data points are shifted up
by a factor $2$ to separate the two data sets.
The two straight lines show the power law fits
$j/N_d\approx 0.014173\,\gamma^{1.4995}$ (symplectic case) and
$j/N_d\approx 0.014207\,\gamma^{1.5016}$ (projected case) for
the range $0.04\le\gamma\le 0.3$.
The statistical error of the exponents is close to $0.2\%$ in both cases. 
{\it The right panel} shows the decay of $\gamma_j(M)$ with $M$ for $j=0$ 
(red crosses), $j=2$ (green open squares) 
for the UPFO for the projected case of  map 
(\ref{eq_separatrixmap}). 
The lower/blue straight line corresponds to the power law fit 
$\gamma_0(M)\approx 2.26\,M^{-1.13}$ and the upper/pink straight 
line to the fit $\gamma_2(M)\approx 1,95\,M^{-0.86}$ 
(for the range $400\le M\le 1600$). The eigenvector corresponding to 
$\gamma_2$ is localized near the two stable islands of the resonance $1/2$. 
}
\end{figure}

In Fig.~\ref{fig10}, in analogy to Fig.~\ref{fig4}, 
we show the dependence 
of integrated number of eigenvalues
$j/N_d$ on $\gamma_j = -2\ln |\lambda_j|$
for the symplectic and projected cases of the UPFO
of the map (\ref{eq_separatrixmap}). In both cases 
we have approximately the same dependence
with the algebraic exponent
$\beta \approx 1.5$ which works for the range
$0.04 \leq \gamma \leq 0.3$.
The minimal values of $\gamma$
(e.g. $\gamma_0$ and $\gamma_2$)
drop approximately inversely proportionally
to $M$. As for symplectic case \cite{frahmulam}
we attribute this decrease with $M$
to a finite size coarse-graining effect of the Ulam method.
As in \cite{frahmulam},
we argue that the exponent $\beta$
for a more physical  intermediate  range of $\gamma$
is directly related to the Poincar\'e exponent. 

\begin{figure}[h]
\begin{center}
\includegraphics[width=0.48\textwidth]{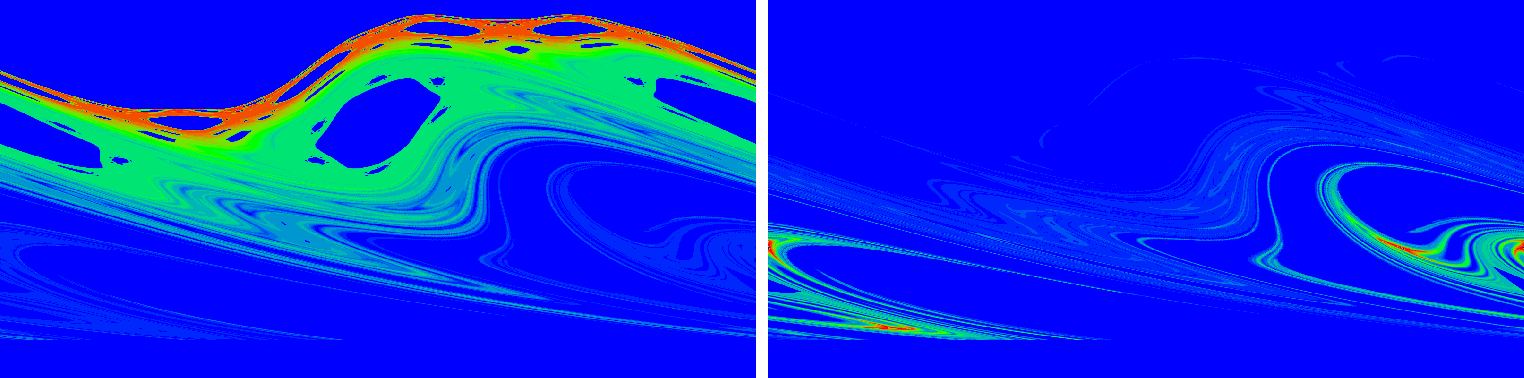}
\end{center}
\caption{\label{fig11} (Color online)
Density plot of the modulus of the eigenvector components of the UPFO for the 
projected case of the map (\ref{eq_separatrixmap})
at $M=1600$ for the two modes with $\lambda_0=0.99972660$
(left panel) and 
$\lambda_{77}=-0.49158775+i\,0.85153885\approx |\lambda_{77}|\,e^{i\,2\pi(1/3)}$
(right panel). 
}
\end{figure}

Examples of two eigenmodes at $\lambda_0$ and 
$\lambda_{77}$ are shown in Fig.~\ref{fig11}.
In the first case we have an eigenmode of diffusive type
similar to Fig.~\ref{fig5} while in the latter case
we have an eigenmode concentrated around unstable 
fix points of resonance $1/3$ 
(see corresponding state of symplectic case
in bottom left panel of Fig.11 in \cite{frahmulam}).

\begin{figure}[h!!]
\begin{center}
\includegraphics[width=0.48\textwidth]{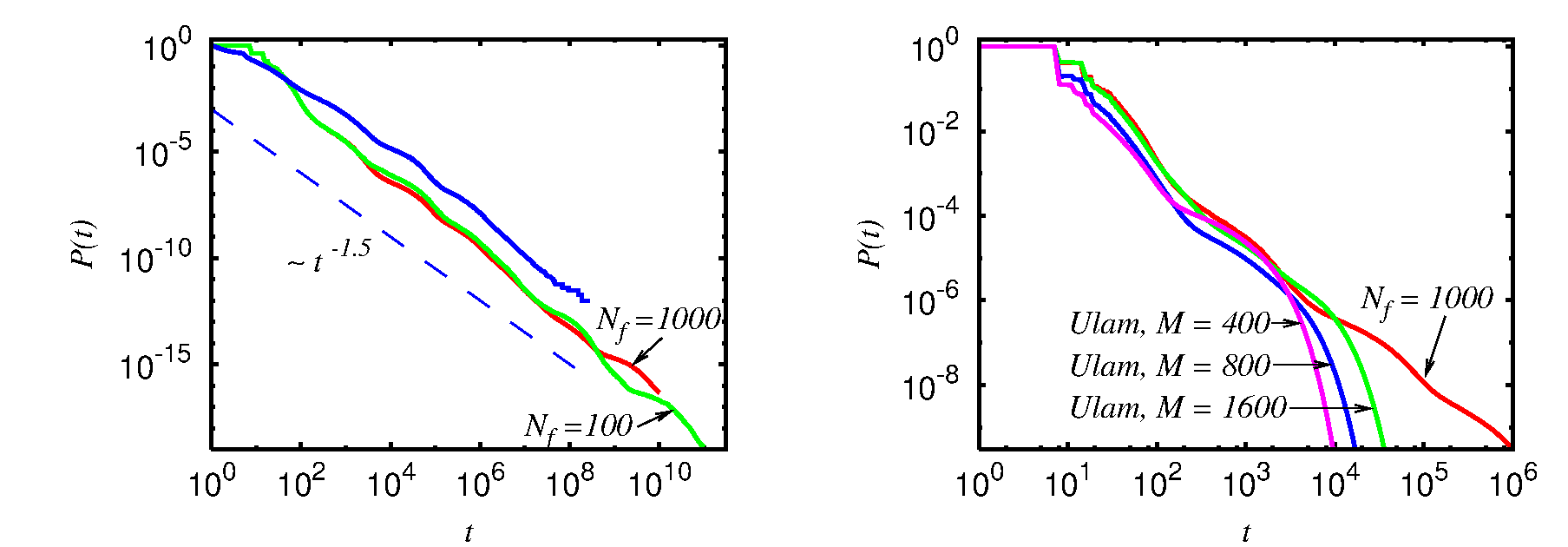}
\end{center}
\caption{\label{fig12} (Color online)
{\it The left panel} shows the average over 10 random realizations 
of the statistics of Poincar\'e recurrences $P(t)$ of the map 
(\ref{eq_separatrixmap}), obtained by the SMCM. 
The red curve, for $t\le 10^{10}$, corresponds to $N_f=1000$. 
The green curve, for $t\le 10^{11}$, corresponds 
to $N_f=100$. The upper/blue curve, for $t\le 2.8\times 10^8$, 
corresponds to the data shown in \cite{chirikov1999b} using a 
direct computation of the statistics of Poincar\'e recurrences. 
{\it The right panel} compares  
$P(t)$  SMCM data for $N_f=1000$ (red curve in left and right panels)
with $P(t)$ obtained by the 
Ulam method for $M=400,\,800,\,1600$. 
At  large times $t>t_{\rm exp}\sim 2\times\,10^{3}-2\times\,10^{4}$ 
the Ulam method leads to an exponential decay  
$P(t)\sim \lambda_0^t$ determined by the largest eigenvalue of the 
UPFO for the projected case.
}
\end{figure}

The comparison of the statistics of Poincar\'e recurrences obtained 
from the map 
(\ref{eq_separatrixmap}) by the SMCM and the usual method are shown
in Fig.~\ref{fig12}. The data of the 
usual method obtained in \cite{chirikov1999b}
allows to follow the decay of $P(t)$ up to $t =2 \times 10^8$,
while with the SMCM we reach times $t=10^{10}$ with
$N_f=1000$ and $t=10^{11}$ with $N_f=100$.
We have a good agreement between three curves
for the range $100 \leq t \leq 10^8$ 
with a certain constant displacement in $\log_{10} t$
of data from the usual method compared to the SMCM data.
This shift appears due to different initial conditions
but apart of this shift all oscillations 
of $P(t)$ curve are well reproduced.
This shows that both methods works correctly. 
However, with the SMCM we are able to reach times 
being by one to two orders of magnitude larger
than previously.

\begin{figure}[h]
\begin{center}
\includegraphics[width=0.48\textwidth]{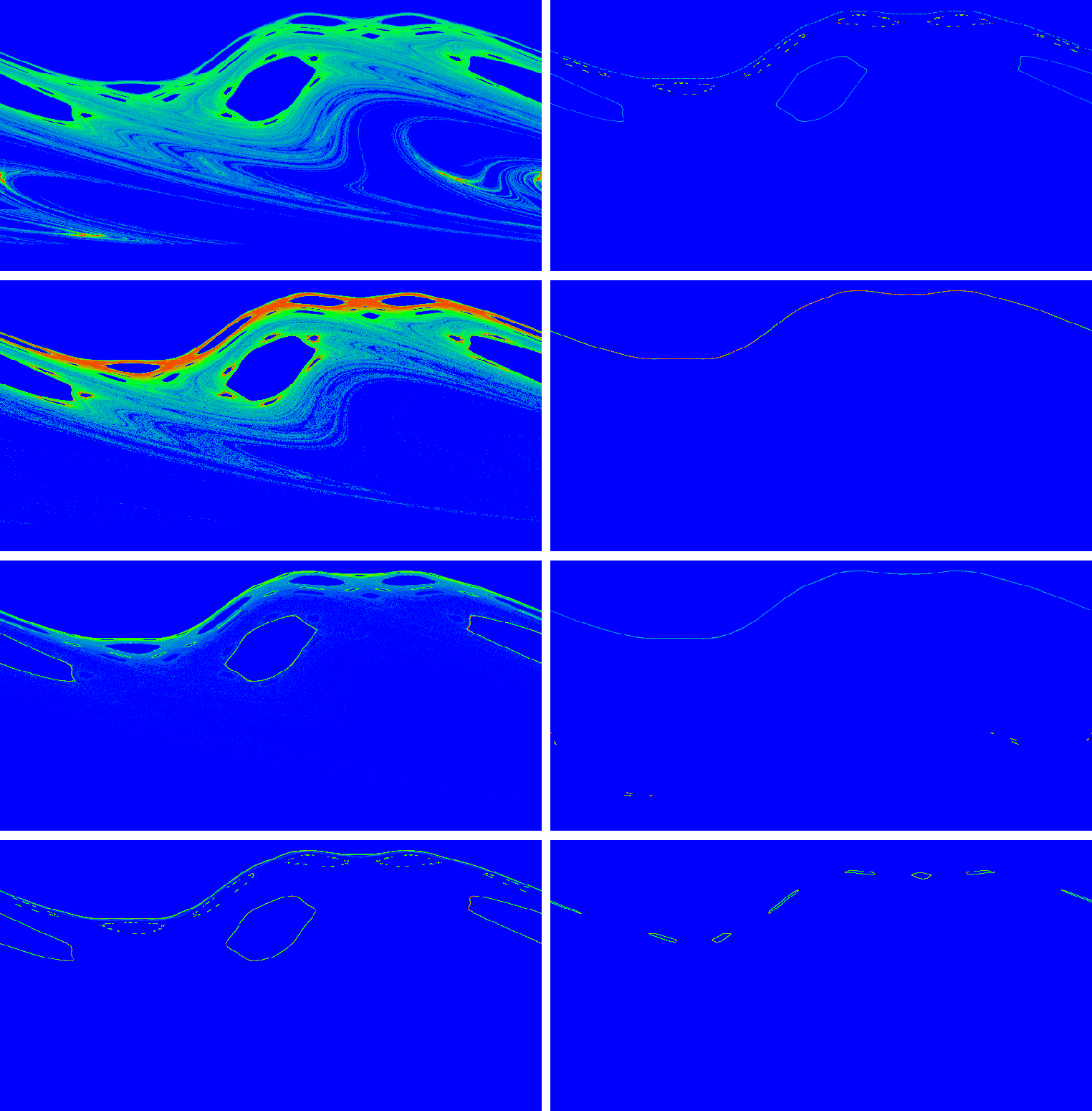}
\end{center}
\caption{\label{fig13} (Color online)
Density plots of the trajectories of the SMCM with $N_f=1000$ 
for the map (\ref{eq_separatrixmap}) 
for various times $t$ and various realizations. All 
density plots are obtained by a histogram of $10^7$ data points 
with a resolution of $800\times 400$ cells for the phase space 
$0\le x< 1$ and $0\le y<4 $. The data points are obtained 
by iterating the $N(t)$ trajectories (with $N(t)=P(t)\,N_i$ for 
$P(t)\ge 10^{-3}$ and $N(t)=N_f$ for $P(t)<10^{-3}$) from $t$ to 
$t+\Delta t$ with $\Delta t=10^7/N(t)$. The left four panels and the upper 
right panel correspond to one particular random realization at  
$t=10^2,\,10^4,\,10^6,\,10^8,\,10^{10}$ and the three lower right 
panels correspond to three other random realizations at $t=10^{10}$. 
For short times $t<10^{5}$ there is no significant difference between the 
density plots for different random realizations at a given time. 
}
\end{figure}

The algebraic fit of SMCM data in Fig.~\ref{fig12}
gives $\beta =1.855 \pm 0.004$ for $N_f=100$
(range $10^4 \leq t \leq 10^{11}$)
and $\beta = 1.706 \pm 0.004$ for
 $N_f=1000$
(range $10^4 \leq t \leq 10^{10}$).
In both cases the statistical error is rather small
but there are visible fluctuations
which become to be significant at $t >10^9$ for 
$N_f=100$ even if they are smaller compared to 
the similar case of map (\ref{eq_stmap})
shown in Fig.~\ref{fig8}.
Due to that one should take as the reliable value
$\beta =1.706$ that shows a noticeable difference
from the value $\beta =1.587$ found above for 
the Chirikov standard map at $K=K_g$.

The comparison of the SMCM data for $P(t)$ with the results
of the Ulam method are shown in the right panel
of Fig.~\ref{fig12}. As it was the case for the similar comparison 
shown in Fig.~\ref{fig8} we find that both methods give the same 
results but the Ulam method works only for time scales being
significantly smaller than those reached with the SMCM.

Finally, as in Fig.~\ref{fig9}, we show in Fig.~\ref{fig13}
the density distribution obtained for various realizations
and various times of the map (\ref{eq_separatrixmap}).
The situation is similar to Fig.~\ref{fig9}:
at short times the density is bounded by cantori barriers,
at large times it reaches the critical golden curve
and at even larger times we see that the density
is located near the critical golden curve or
other secondary resonances depending on the realization.

\section{Properties of eigenstates of Ulam matrix}

\begin{figure}[h]
\begin{center}
\includegraphics[width=0.48\textwidth]{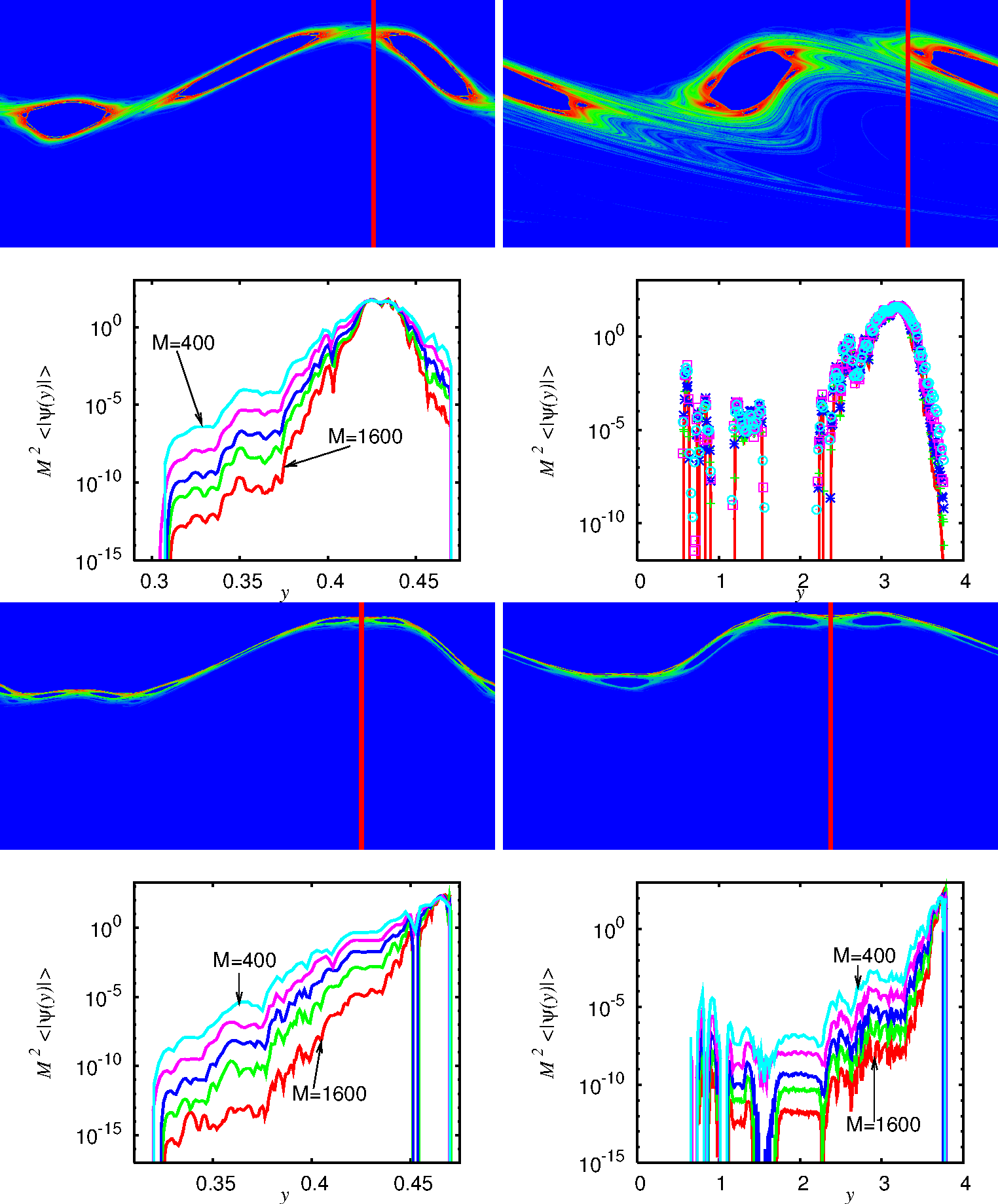}
\end{center}
\caption{\label{fig14} 
(Color online) The localization properties in $y$-direction for certain 
eigenvectors of the UPFO for the projected case for the maps 
(\ref{eq_stmap}) (left column) and (\ref{eq_separatrixmap}) (right column).
The panels in the second and fourth row show the averaged modulus 
$\langle |\psi(y)|\rangle$ of the eigenvector components within 
a band of  $1$\% width of the whole $x$-range at a certain $x=x_0$. 
The global structure of the corresponding eigenstates
is shown in the corresponding first and third panels 
(counting from the top; the red vertical thick line indicates the range of 
$x$-values where the average has been performed for each $y$-value,
$M=1600$). 
Data are shown for $M=400$ (cyan/highest curve), $M=560$ 
(pink/second curve), $M=800$ (blue/third curve), $M=1120$ (green/fourth curve)
and $M=1600$ (red/lowest curve). In the right panel of the second row 
the data for different values of $M$ approximately coincide 
and only the data for $M=1600$ are shown by a full (red) curve;  other 
$M$ values are shown as isolated data points for 
$M=1120$ (green crosses), $M=800$ (blue stars), $M=560$ (pink squares) 
and $M=400$ (cyan circles). 
For $M=1600$ the eigenvectors, shown in the density plots of the first and 
third row, correspond to the modes $\lambda_4$ and $\lambda_{31}$ 
of the map (\ref{eq_stmap}) (left column) and to the modes $\lambda_2$ 
and $\lambda_{17}$ of the map (\ref{eq_separatrixmap}) (right column);
for other $M$ we show corresponding eigenvector located at the same resonances.
}
\end{figure}

Let us now try to analyze how the decay of Poincar\'e recurrences
is related to the properties of the (right) eigenvectors $\psi(x,y)$ 
of the UPFO for the projected case. 
For this we determine the $x$-average of the eigenvector amplitude 
around a given position $x_0$ 
over a band of $1$\% width of the whole $x$-range:
$\langle |\psi(y)|\rangle=100\,M^{-1}\,\sum_{|\Delta x|<0.005} 
|\psi(x_0+\Delta x,y)|$. The $y$-dependance of this average allows 
to visualize the localization properties 
of the eigenstate in $y$-direction. In Fig.~\ref{fig14} we show 
this quantity for two examples for each of the maps 
(\ref{eq_stmap}) and (\ref{eq_separatrixmap}) and for different values 
of $M$ between 400 and 1600. 

For the case of the map (\ref{eq_stmap}),
shown in the left column of Fig.~\ref{fig14},
we see a clear evidence of exponential localization of
eigenstates. In fact the average amplitude 
in a vicinity of $y \approx 0$, where the initial state is
taken and where the absorption happens,
has enormously small values being of the order of $10^{-15}$.
These amplitudes on the tail drop significantly with an increase
of $M$. For the map (\ref{eq_separatrixmap})
the decay of eigenstates is more irregular since the
band at $x\approx x_0$ crosses some secondary islands
thus leading to appearance of a plateau in the decay with $y$.
But in global we can still say that there is an exponential
decay of eigenstates. This exponential localization 
of eigenstates reminds the Anderson localization
in disordered solid state systems 
(see e.g. \cite{anderson}).

\begin{figure}[h]
\begin{center}
\includegraphics[width=0.48\textwidth]{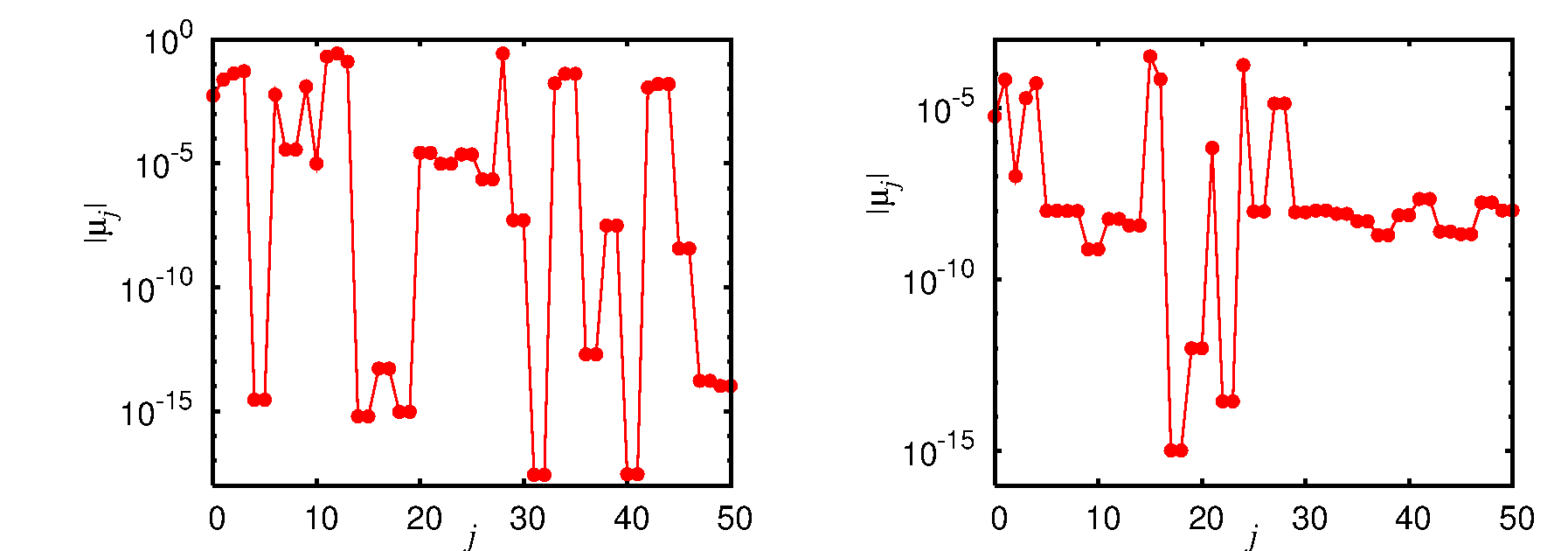}
\end{center}
\caption{\label{fig15} 
(Color online) Modulus of the projection coefficients $\mu_j$ of the 
initial density vector $\psi_{\rm init}$, localized in one cell at 
$x_0=y_0=0.0625$, with respect to the right eigenvectors $\psi_j^R$ (of 
the UPFO projected case for $M=1600$) 
versus level number $j$. These 
coefficients appear in the expansion $\psi_{\rm init}=\sum_j \mu_j\,\psi_j^R$
(see text).
{\it The left and right panels} represent data 
for  the maps (\ref{eq_stmap}) and 
(\ref{eq_separatrixmap}) respectively. The 
cases with $|\mu_j|=|\mu_{j+1}|$ correspond to pairs of complex conjugated 
modes with $\mu_{j+1}=\mu_j^*$. 
}
\end{figure}

We can also consider the projection of our 
initial state taken in a cell $\ell_0$ on the eigenstates. 
Indeed, this initial state can be expressed as 
$\psi_{\rm init}=\sum_j \mu_j\,\psi_j^R$
where $\mu_j$ are expansion amplitudes and
$\psi_j^R$ the right eigenvectors defined by Eq.~(\ref{eq_eigen}).
To determine the values of $\mu_j$ we need first to compute
the left eigenvectors $\psi_j^L$ of the Ulam matrix $S_p$
which are biorthogonal to the right eigenvectors $\psi_j^R$ and 
provide the expansion amplitudes by the identity:
$\mu_j=\<\psi_j^L|\psi_{\rm init}\>/\<\psi_j^L|\psi_j^R\>$. 
Note that this expression does not depend on the choosen normalization 
of the eigenvectors and it requires only that $<\psi_j^L|\psi_j^R\>\neq 0$. 
However, for convenience, we have normalized both type of eigenvectors by the 
$L_1$-norm such that $\sum_{x,y} |\psi_j^{R,L}(x,y)|=1$. 
We have numerically determined the first 51 left eigenvectors with the help of 
the Arnoldi method applied to the transpose of $S_p$ and therefore obtained 
the corresponding expansion amplitudes. 

\begin{figure}[h]
\begin{center}
\includegraphics[width=0.48\textwidth]{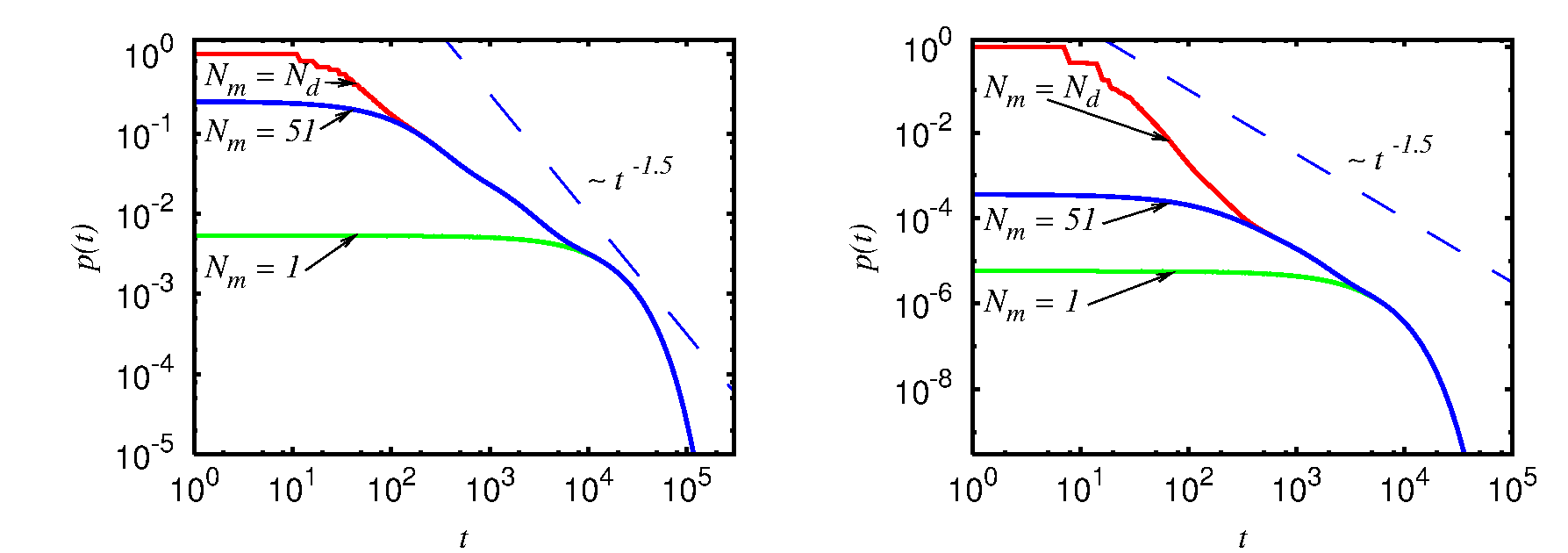}
\end{center}
\caption{\label{fig16} 
(Color online) Contributions of the largest eigenmodes 
of the UPFO projected case at $M=1600$ 
to the statistics of Poincar\'e recurrences for the maps 
(\ref{eq_stmap}) (left panel) and (\ref{eq_separatrixmap}) (right panel).
Here, we show the probability $p(t)$ obtained 
from the expansion over eigenvectors given by 
the formula $p(t)=\sum_{j=0}^{N_m-1} p_j\,\lambda_j^t$
with $p_j=\mu_j\sum_{x,y}\psi_j^R(x,y)$, $N_m$ being the number of used 
modes and the eigenvalues being ordered as 
$|\lambda_0|>|\lambda_1|>|\lambda_2|>\cdots$ (see text).
The upper red curve is obtained from the direct iteration of 
the UPFO (see green curve in the right panels of figures \ref{fig8} and 
\ref{fig12}) and corresponds to the contribution of the full spectrum of 
all eigenvalues with $N_m=N_p$. The 
middle blue curve corresponds to $N_m=51$ with the same $\mu_j$ values as 
those shown in figure \ref{fig15}. The main contributions to this curve 
arise from the diffusion modes (with real positive eigenvalues $\lambda_j>0$),
the other resonant modes with complex or real negative 
eigenvalues give only a small contribution which does not modify 
the curve up to graphical precision. 
The bottom green curve corresponds to $N_m=1$, 
i.~e. the contribution $\mu_0\,\lambda_0^t$ of the largest $\lambda$ 
eigenmode. In both panels the dashed line indicates for comparison
a power law decay $P(t) \propto t^{-1.5}$.
}
\end{figure}

The dependence of $\mu_j$ on $j$ is shown in Fig.~\ref{fig15}.
We see that there are enormously large
fluctuations of $\mu_j$ which are in a range of $10$ orders of magnitude. In 
particular the amplitudes corresponding to resonant modes are very small 
which is easy to understand if the resonant mode is localized far away from 
the initial state and does therefore not contribute to the expansion. 
We think that these fluctuations are at the origin
of the slow algebraic decay of Poincar\'e recurrences $P(t)$ (see below).

In Fig.~\ref{fig16} we show the contribution of the 
largest $N_m$ eigenmodes to the statistics of Poincar\'e recurrences 
(for $M=1600$) given by the formula: 
$p(t)=\sum_{j=0}^{N_m-1} p_j\,\lambda_j^t \;\;$
with $\;\; p_j=\mu_j\sum_{x,y}\psi_j^R(x,y)$ and the eigenvalues 
ordered as $|\lambda_0|>|\lambda_1|>|\lambda_2|>\cdots$. 

For $N_m=N_p$, we have the statistics of Poincar\'e recurrences 
obtained from the iteration of the UPFO and already shown in 
Figs.~\ref{fig8} and \ref{fig12}. For $N_m=51$ we have evaluated 
the sum using the expansion coefficients shown in Fig.~\ref{fig15}. 
Both curves coincide at $t>10^2$ for the map (\ref{eq_stmap}) 
or at $t>3\times 10^2$ for the map (\ref{eq_separatrixmap}) showing that 
the largest eigenmodes determine 
the long time behavior. For large times ($t> 10^4-10^5$) only the 
first eigenmode contributes and the decay is purely exponential. 
It turns out that in the sum for $N_m=51$ the terms 
arising from the resonant modes can be omitted without changing the curve 
up to graphical precision since these modes contribute only very weakly in 
the expansion. 
In general, the partial sum $p(t)$ converges to the actual statistics of 
Poincar\'e recurrences $P(t)$ with increasing $N_m$ and at 
given value of $N_m$ one expects that $p(t)$ and 
$P(t)$ coincide for $t\gg 2\,\gamma_{N_m}^{-1}$. 

The data of Figs.~\ref{fig14},~\ref{fig15},~\ref{fig16} illustrate 
the nontrivial link between the localized eigenstates
of the Ulam matrix and the decay of Poincar\'e recurrences.
The eigenmodes are exponentially localized and 
for many of them their projection on the initial state is very small
but at some large times their contribution can become
very important since the modes with large projections
decay more rapidly.

\section{Discussion}

Our studies show that the generalized Ulam method
reproduces well the decay of Poincar\'e recurrences $P(t)$
in 2D symplectic maps with divided phase space.
At the same time the computation of $P(t)$
is obtained in a more efficient way by
the proposed SMCM allowing to reach time scales
of the order of $t=10^{10}$. We find that at these large
times the Poincar\'e exponent has values $\beta =1.58$
for the Chirikov standard map at $K_g$
and $\beta =1.70$ for the separatrix map
at $\Lambda_c$. The recurrences at large times are
dominated by sticking of trajectories
not only in a vicinity of the critical golden curve
but also in a vicinity of secondary resonance structures.
This confirms earlier numerical observations
obtained on shorter time scales \cite{ketzmerickpre}.

The sticking around various different resonant structures
on smaller and smaller scales of phase space
leads to nontrivial oscillations of
the Poincar\'e exponent. The values of $\beta$
found here are not so far from the average
values found previously 
by averaging over maps at different parameters with
$\beta \approx 1.5$ \cite{kiev,kievb},
 $\beta \approx 1.57 $ \cite{ketzmerick}.
In agreement with the
data presented here and in \cite{frahmulam}, 
we find that the above value of $\beta$
is close to the exponent of 
integrated density of states
of the Ulam matrix which has $\beta \approx 1.5$.
At the same time we see that at $t=10^{10}$ 
the fluctuations in the Chirikov standard map
at various $N_f$ and various random realizations
are significantly stronger as compared to the  
separatrix map.  

We attribute these fluctuations
to a localization of eigenstates of the Ulam matrix
which gives very nontrivial properties of 
eigenstates projection on an initial state.
The properties of these eigenstates are still poorly
understood. We think that the further developments
of analytical models
of renormalization on Cayley type tree
\cite{ott,ketzmerick,nechaev,agam} 
and their applications to the 
puzzle of statistics of Poincar\'e recurrences 
should develop a more detailed
analysis of localization of
eigenstates of the Ulam matrix.

This research is supported in part by the EC FET Open project 
``New tools and algorithms for directed network analysis''
(NADINE $No$ 288956). This work was granted access to the HPC resources of 
CALMIP (Toulouse) under the allocation 2012-P0110.
We also acknowledge the France-Armenia collaboration grant 
CNRS/SCS $No$ 24943 (IE-017) on ``Classical and quantum chaos''.

We dedicate this work  
to the memory of Boris Chirikov
(06.06.1928 - 12.02.2008).


\end{document}